\documentclass{aa}
\input epsf.tex
\begin{document}

\thesaurus{10.05.1;10.06.1;10.07.2;10.08.1}

\title{The Old Halo metallicity gradient: \\
the trace of a self-enrichment process}

\author{G. Parmentier, 
        E. Jehin, 
        P. Magain, 
        A. Noels and 
        A.A. Thoul\thanks{Chercheur qualifi\'e au 
                    Fonds National de la Recherche Scientifique (Belgium)}}   
\offprints{G.\ Parmentier}

\institute{Institut d'Astrophysique et de G\'eophysique, Universit\'e de 
           Li\`ege, 5, Avenue de Cointe, B-4000 Li\`ege, Belgium}

\date{Received date; accepted date}

\maketitle
\markboth{G.\ Parmentier et al.: The Old Halo metallicity gradient}{} 

\begin{abstract}

Based on a model of globular cluster self-enrichment 
published in a 
previous paper, we present an explanation for the metallicity gradient
observed throughout the galactic Old Halo.
Our self-enrichment model is based on the ability of globular cluster
progenitor clouds to retain the ejecta of a first generation of Type II
Supernovae.  The key point is that this ability depends on the pressure 
exerted on the progenitor cloud by the surrounding protogalactic medium
and therefore on the location of the cloud in the protoGalaxy.
Since there is no significant (if any) metallicity gradient in the whole halo,
we also present a review in favour of a galactic halo partly build 
via accretions and mergers of satellite systems.  Some of them bear 
their own globular clusters and therefore ``contaminate'' 
the system of globular clusters formed ``in situ'', namely 
within the original potential well of the Galaxy.  
Therefore, the comparison between our 
self-enrichment model and the observational data should be limited to 
the genuine galactic globular clusters, 
the so-called Old Halo group.

\keywords{Galaxy: evolution -- Galaxy: formation -- 
globular clusters: general -- Galaxy: halo}
\end{abstract}

\section{Introduction}
\label{sec:intro}
  
Galactic globular clusters (hereafter GCs) 
are fossil records of the formation of the Galaxy.
The understanding of their formation process would certainly shed light 
on the early galactic evolution.  However,  
at the present time, there is no widely accepted theory of GC formation.  
In Parmentier et al. (1999) (hereafter Paper I), we suggest a 
formation scenario based on a self-enrichment process 
such as proposed by Cayrel (1986)
and further developed by Brown et al. (1991, 1995).

Our self-enrichment scenario takes place within the Fall \& Rees (1985) 
description of the protoGalaxy, namely cold clouds embedded in a 
hot protogalactic background.
These cold clouds are assumed 
to be the progenitors of galactic halo GCs.
Since they are made up of primordial gas,   
the main advantage of a self-enrichment scenario is that 
it explains in parallel the formation of the clusters and the origin 
of their metal contents.  \\
\indent The main target of Paper I was to 
demonstrate conclusively that the gaseous 
progenitors of galactic halo GCs are able 
to sustain a few hundreds of Type II Supernovae (hereafter SNII)
without being disrupted.  
This result is in contrast with the widespread idea according to which
a few supernovae are able to disrupt a Proto-Globular Cluster Cloud 
(hereafter PGCC).
Furthermore, the large number of SNeII allowed by our model 
can explain the amount 
of metals currently observed in galactic halo globular clusters,
and this without any requirement of pre-enrichment of the gas. \\
\indent The aim of the present paper is to explore further an 
interesting consequence of Paper I, which is also the main 
difference existing between our self-enrichment model and the one 
developed by Brown et al.~(1995).
The metallicity that a PGCC can reach through 
self-enrichment depends on 
the pressure exerted by the medium surrounding the progenitor cloud 
and, therefore, on the cloud location in the protoGalaxy.
The deepest in the protoGalaxy the PGCC is located, the highest 
the final metallicity induced by self-enrichment will be.
Therefore, we expect to find a metallicity gradient throughout 
the galactic halo. 

The paper is organised as follows. \\
In Sect.\,2, we briefly review the self-enrichment model
presented in Paper I, focusing on the link 
between the final metallicity of the PGCC and the pressure exerted 
on it by the surrounding hot protogalactic background. 
In Sect.\,3, we examine the 
different arguments suggesting the existence of 
halo substructures, in order to isolate 
to which one the self-enrichment model can be safely compared.
In Sect.\,4, we compare the model with 
the observations.  
Sect.\,5 explores the putative link between GC metallicities and 
their perigalactic distances.
Finally, we present our conclusions in Sect.\,6.

\section{Self-enrichment model}
\indent According to Fall \& Rees (1985), galactic halo GCs were formed  
during the collapse of the protoGalaxy.
During this collapse, a thermal instability triggers the development of 
a two-phase structure, namely 
cold and dense clouds in pressure equilibrium with a hot and diffuse 
protogalactic background.  
The temperature of the clouds is assumed to remain at a value of 10$^4$\,K,
where the cooling rate drops sharply in a metal-free medium.
Their masses scale as the Jeans mass of a pressure-truncated 
spherical cloud with a temperature
$\simeq 10^4$\,K. 
Since this is of the same 
order of magnitude ($\simeq$ 10$^6$\,M$_{\odot}$) as the GC masses, 
Fall \& Rees (1985) identify these cold clouds 
with the progenitor clouds of GCs (however, this temperature, 
and therefore, the characteristic mass, is preserved only if there 
is a UV flux able to prevent any H$_2$ formation, the 
main coolant in a metal-free gas below 10$^4$\,K).
As already mentioned, the PGCCs are assumed 
to be metal-free and the 
formation process must therefore explain how the metals are provided within 
each cloud. 
Within this context, the self-enrichment hypothesis was proposed by 
Cayrel (1986) and further developed by Brown et al.~(1991, 1995).
A first generation of stars is assumed to form in the central regions 
of the progenitor cloud.  When the massive stars of this first 
generation explode as SNeII, 
all the cloud material is progressively swept in an expanding supershell.
This supershell gets chemically enriched with the metals released 
by the exploding massive stars.  Since it is a 
compressed layer of gas, it constitutes  a dense medium where 
the formation of a second generation of stars is triggered.  
Under favourable conditions (see Brown et al., 1995) 
these second generation stars, formed in the chemically
enriched supershell, can recollapse and form a GC.
Therefore, the first generation SNeII provide the GC metals and 
trigger the formation of the GC stars. \\ 
\indent Supernova energetics has been a major criticism of
the GC self-enrichment hypothesis.
However, the main target of Paper I was 
to conclusively demonstrate that a PGCC is not 
necessarily disrupted by SNII explosions.
Our self-enrichment model is detailed in Paper~I.
Suffice it to say that we compute, for a given hot protogalactic 
background pressure, the supershell 
velocity during the sweeping of the PGCC as a function of the explosion 
rate.  Based on this 
result, we compare the kinetic energy of the supershell with 
the binding energy of the PGCC in order to get the maximum number of SNeII
the cloud is able to sustain. \\
\indent Whatever the value of $P_{h}$, 
the pressure exerted by the hot protogalactic background 
on the cloud, we find that a PGCC can sustain about 200 SNeII. 
Such a large number of SNeII
can provide the amount of metals observed in galactic halo GCs. 
The results of Paper~I are summarized in Table~\ref{tab} (see Sect.~4.2 
for a justification of the $P_h$ values). 
Clearly, GC halo metallicities can be reached through self-enrichment.

\begin{table}[t]
  \caption{PGCC masses and metallicities for different values of the 
   pressure of the medium confining the PGCCs.}
  \label{tab}
  \begin{center}
    \leavevmode
    \footnotesize
    \begin{tabular}[h]{ccc}
      \hline \\[-5pt]
      $P_{\rm h}$ [dyne.cm$^{-2}$] & log$_{10}$ M/M$_{\odot}$  
               &  [Fe/H] \\[+5pt]
      \hline \\[-5pt]
      10$^{-11}$  & 6.5 & -2.2  \\
      10$^{-10}$  & 6.0 & -1.7  \\
      10$^{-9}$   & 5.5 & -1.2  \\   
      \hline \\
      \end{tabular}
  \end{center}
\vspace{-0.7 cm}
\end{table}

We see from Table~\ref{tab} that
the higher the external pressure is, the higher the metallicity will be.
Indeed, since the dynamical constraint leads to
a constant SNII number, assuming a given Initial Mass Function 
(namely a Salpeter 
one) and a given stellar mass range, a constant amount of metals 
(independent of $P_h$) is 
released by the first generation massive stars.
Since the PGCC mass decreases with increasing external pressure
(the Jeans mass scales as $P_{h}^{-1/2}$, 
Eq.~(5) in Paper~I), 
the PGCCs embedded in a higher pressure medium, namely
located deeper in the protoGalaxy, reach higher final metallicities.
This self-enrichment model, contrary to the one developed by 
Brown et al.~(1995),
implies a {\it metallicity gradient} throughout the galactic halo. \\  
\begin{figure}[t]
  \begin{center}
    \leavevmode
    \epsfxsize= 8.5 cm 
    \epsffile{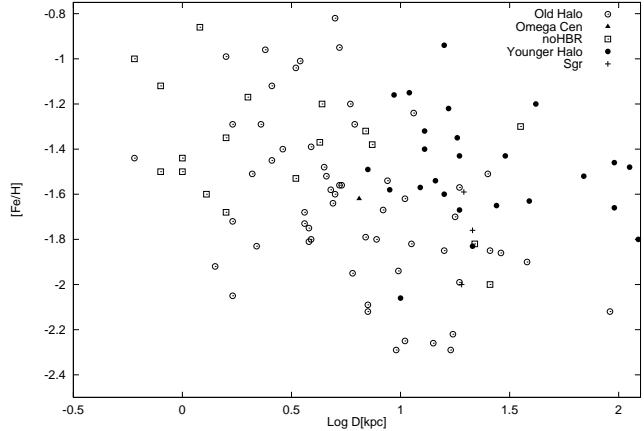}
  \end{center}
\caption{The whole galactic Halo exhibits no clear-cut 
dependence of the metallicity on the galactocentric distance. 
It includes different subpopulations, 
such as the Old Halo and the Younger Halo.  
For the meaning of the different symbols, see Sect.\,4.1.
Data are based on Harris 1996\label{FeHDWH}}
\end{figure}
At first sight, there is no confirmation of this self-enrichment model by
the observational data.  The galactic halo exhibits no significant
metallicity gradient (see Fig.~\ref{FeHDWH}; data are based on Harris 1996
\footnote{Data updated 1999 June 22 at 
http://physun.physics. mcmaster.ca/Globular.html.})
However, according to Zinn (1993), the galactic halo is
composed of {\it two distinct subpopulations of GCs}~, what he calls
an Old Halo and a Younger Halo.  
The next section presents a review of the evidence supporting Zinn's idea.  

\section{Two populations of galactic halo GCs}
\label{sec:twopop}

According to Zinn (1985), the galactic Globular Cluster System 
(hereafter GCS) includes two subsystems: the 
disk GCs \footnote{Due to their high concentration near the 
galactic center, these metal-rich clusters, at least those located 
within the inner $\sim$4\,kpc of the Galaxy, are now thought to 
be associated with the galactic bulge
rather than with the disk (Minnitt 1995, C\^ot\'e 1999).} ([Fe/H] $>$ 
$-$0.8) and the halo GCs ([Fe/H] $<$ $-$0.8).  
Both groups also differ in their mean rotational velocity and 
spatial distribution about the galactic center.  \\
Globular cluster classification has now gone a step further, and 
the halo subsystem could be itself divided into two groups.

\subsection{Horizontal Branch morphology}

Zinn (1993) sorts the galactic halo GCs
into two groups according to their Horizontal Branch
(hereafter HB) morphology.  
The HB morphology, namely the colour distribution of the 
stars located along the HB, can be described by the 
index $C=(B-R)/(B+V+R)$ where B, V and R are respectively the numbers 
of blue, variables and red HB stars.  This index therefore ranges from $-$1
for a purely red HB to +1 for a purely blue one. 
The morphology of the HB is essentially driven by the metallicity 
of the cluster.  As such, [Fe/H]  is the {\it first parameter} that governs 
the HB morphology.  However, it has long been known that some clusters
 with similar [Fe/H] values present very different HB morphologies
(e.g. M\,13 and M\,3, NGC\,288 and NGC\,362).
 This is the so-called {\it second parameter} effect: a 
second parameter (at least), in addition to metal abundance,
is needed to explain the HB morphology. \\

Zinn (1993) notices that, 
while the inner halo GCs ($D <$ 8\,kpc, where $D$ 
is the galactocentric distance) exhibit a 
tight relationship between HB morphology and [Fe/H],
the outer halo GCs ($D >$ 8\,kpc) show a large scatter 
in the same relation.  Hence, the requirement of a 
second parameter to explain the HB morphology is 
mostly needed for GCs located beyond the solar circle.
Since, at a given metallicity, these clusters have also 
redder HBs than the inner halo GCs,
Zinn (1993) divides the galactic halo GCs into two groups:
the inner halo GCs are classified as Blue Horizontal Branch (BHB) clusters,
while the clusters whose HB types are 0.4 redder than 
their inner halo counterparts (equal metallicity) are labelled as Red 
Horizontal Branch (RHB) clusters. As mentioned above, 
these RHB clusters are mostly located outside the solar circle. 
This link between the influence of the second parameter and the 
galactocentric distance indicates that the second parameter problem must be 
related in some way to the formation of the Galaxy (Searle \& Zinn 1978). \\ 
\indent According to Lee et al.\,(1993), the most promising candidate for 
this second parameter is age.  {\it If} age is indeed the second 
parameter, then, at a given metallicity, a RHB GC 
is younger than its BHB counterpart.
Therefore, Zinn (1993) labels the BHB group  and 
the RHB group  respectively ``Old'' Halo and ``Younger'' 
Halo. 
To explain these putative age differences, 
and since the ``Younger'' Halo GCs are mostly located in the outer part of 
the halo, Zinn (1993) suggests that the two groups were formed in two 
different ways.
The ``Old'' Halo (hereafter OH) GCs would have been formed during 
the rapid collapse of the protogalactic cloud, while the ``Younger'' 
Halo (hereafter YH) GCs would have been 
formed in satellite systems, such as dwarf galaxies, 
which evaded the protogalactic collapse 
and which were later accreted by the Galaxy.

\subsection{GC Ages}

The hypothesis that RHB clusters are younger than their BHB counterparts 
relies on the implicit assumption that age is the only 
second parameter driving the HB morphology, in addition to metallicity.
Therefore, it is interesting to wonder if age determinations based 
on the luminosity or the colour of the Main Sequence Turn-Off confirm
such an hypothesis.  \\
\indent The nature of the second parameter remains a much debated question
(for contrasting points of view, see the reviews by Stetson et al.\,1996
and Sarajedini et al.\,1997). 
Chaboyer et al. (1996) find that RHB clusters 
are on average 2-3\,Gyr younger than BHB clusters, which is consistent
with age being the dominant second parameter.  \\
This point of view has been reinforced 
by other studies where some clusters have been found 
to be significantly younger than the bulk of the other galactic halo GCs.  
These are Pal\,12 (Stetson et al.\,1989), 
Rup\,106 (Buonanno et al.\,1993), Arp\,2 (Buonanno et al.\,1995), 
IC\,4499 (Ferraro et al.\,1995), Pyxis (Sarajedini \& Geisler 1996),
Pal\,14 (Sarajedini 1997),
Pal\,5 (Buonanno et al.\,1998), the outer halo GCs Pal\,3, Pal\,4, 
Eridanus (Stetson et al.\,1999), NGC\,362, 1261, 1851, 2808 
(Rosenberg et al.\,1999).
All these younger GCs are RHB clusters and are located
outside the solar circle, where the second parameter effect is dominant.
Zinn's idea is therefore strengthened. 
However, according to Rosenberg et al.\,(1999),
three RHB clusters (NGC\,3201, NGC\,5272 and possibly NGC\,4590) 
have not been proven to
be younger. 
As a result, age appears to be one of the most appealing 
second parameter candidates
(Ferarro et al.\,1995, Chaboyer et al.\,1996, Stetson et al.\,1999),
but an additional parameter is probably at work in a minority of clusters.

\subsection{Kinematic Differences}
 
Whatever the second parameter is, the concept of two genuinely 
distinct halo subsystems gets some support from the presence 
of a kinematic difference between the YH and the OH groups.
While the mean galactic rotation of the OH group is clearly prograde, 
it is close to zero, perhaps even sligthly retrograde,  
for the YH group (Zinn 1993, Da Costa \& Armandroff 1995, 
Odenkirchen et al.\,1997). 
This result has been recently confirmed by the work performed 
by Dinescu et al.\,(1999).
Based on the most complete compilation of GC proper motions 
existing up to now, they compute the 
orbits of 38 GCs.  
Although a sharp truncation between the 
different orbital parameters does not appear,
they show that, on the average, the ``Old'' Halo and the ``Younger'' Halo 
GCs exhibit some differences in their kinematics and their orbit shapes.
The ``Younger'' Halo (RHB) group presents, {\it on the average}, 
a smaller rotational velocity, larger velocity dispersions, 
higher orbital energies, higher apogalactic distances ($D_{\rm a} 
\geq$ 10\,kpc) and higher excentricities than the ``Old'' Halo (BHB) group.
If the RHB group represents an accreted component of our Galaxy, 
the mean rotational velocity suggests that a significant 
fraction of the outer GCs came from one or more 
ancestral objects on retrograde orbits.

\subsection{Dwarf galaxies accretion}

Dwarf irregular and dwarf 
spheroidal galaxies, at least the most massive ones, also host their own GCS
(e.g. the Magellanic Clouds, Fornax). 
Interestingly, in a plot of [Fe/H] vs C, 
the HB morphology index, some of the Large Magellanic Cloud (hereafter LMC)
GCs fall among the outer halo GCs (Da Costa 1993).
{\it If} age is accepted as the dominant second parameter, 
then these clusters are younger than the inner galactic halo GCs.  
This is in agreement with the presence in the LMC of young 
clusters whose masses are within the galactic GC mass range
(Elson \& Fall 1988, Meylan \& Heggie 1997). \\
The hypothesis that the halo was partly built 
via accretion was underlined by several authors.
Lin \& Richer (1992) suggest that Rup\,106 and Pal\,12, 
two RHB clusters known to be younger than other GCs with 
similar metallicity (see Sect.\,3.2), 
were tidally captured by the Galaxy from the Magellanic Clouds
during their recent perigalacticon passage.  
A similar argument holds for Pyxis, another RHB and young cluster
(Irwin et al.\,1995). 
The association between some RHB/younger GCs and streams
(alignments along great circles over the sky
which could arise from the disruption of MW satellites)
is advocated in Majewski (1994) and Fusi Pecci et al.\,(1995). \\
\indent All these GCs could therefore have been born well apart from 
the original protoGalaxy, joining our Galaxy through later infall events.
As such, they are {\it not indicative of the early formation of the 
galactic halo}. \\
   
Furthermore, Nature currently provides us with an example of 
satellite accretion.
The Sagittarius dwarf spheroidal galaxy, the closest satellite of 
the Galaxy, is currently undergoing strong tidal distorsions 
indicating that it will probably be disrupted 
and absorbed by the Milky Way (Ibata et al.\,1997, Johnston et al.\,1999). 
Based on both positional and kinematic data, 4 GCs (M\,54, Arp\,2, 
Ter\,8 with halo metallicities and Ter\,7 with disk metallicity) 
unambigously belong to the Sgr dwarf (Ibata et al.\,1997).
These GCs are therefore being incorporated into the galactic halo
and as such constitute a source of ``contamination'' of the genuine 
galactic GCs, the real tracers of the early evolution of the Galaxy. \\
\indent Ibata et al.\,(1997) also strongly suspect the presence 
of a dark halo around the Sgr dwarf.
Indeed, dwarf spheroidals are among the most dark-matter dominated 
systems known (Mateo 1998) and dwarf irregulars are significanly 
more dark-matter dominated than are large spirals (Carignan et al.\,1990).
The presence of dark 
matter halos around the Milky Way satellites, denser than what is found 
around large spirals, could induce some differences in the star and GC 
formation mechanisms compared to what occurs in the protoGalaxy
(Larson 1993).

\subsection{Spatial distribution}
Finally, considering the GCs with [Fe/H]
$< -$1 (in order to remove the obvious disk clusters),
Hartwick (1987) notes that their spatial distribution can be described 
in terms of two subsystems: an inner flattened distribution and an outer 
more spherical distribution.

\subsection{What does this all mean ?}
  
The convergence of all the differences mentioned above (HB morphologies,
ages, kinematic data, 
galactocentric distances and 
spatial distributions around the galactic center) 
between BHB and RHB GCs adds weight to 
the claim that they form two genuinely distinct groups.
The existence of two main substructures in the galactic halo implies 
that a hybrid picture could conveniently describe its formation
(Stetson et al. 1996, Sarajedini et al. 1997, Rosenberg et al. 1999).
The inner part, populated by BHB GCs, would have been formed 
over a relatively short period of time during the collapse of the
protoGalaxy (Eggen, Lynden-Bell \& Sandage 1962), 
while the outer part, which includes most of the RHB GCs, was mainly 
built via accretion and mergers of 
satellite systems in a still ongoing process 
(Searle \& Zinn 1978).   
In this case, the outer halo objects would actually bear little direct 
relevance to the formation history of the main part of the Galaxy.
As such, they should not be considered when comparing our
self-enrichment model to the observational situation (see Sect.\,4.1). \\ 
\indent To disentangle the {\it genuine} galactic GCs from those formed in 
satellite systems and accreted afterwards, Zinn (1993) suggests 
to rely on a HB morphology criterion.  As illustrated above,  
this approach is indeed fruitful since many RHB 
clusters exhibit peculiarities, such as lower ages than their 
BHB counterparts.  However, the GCSs 
of dwarf galaxies are not exclusively composed of young GCs
and, consequently, the actual situation is certainly more complicated.  
A LMC cluster, Hodge\,11, is as old as the inner halo GCs
(Mighell et al.\,1996).  
The Sagittarius cluster system will contribute to 
both the YH (Ter\,7 and Arp\,2) and OH (Ter\,8 and M\,54) groups
(Da Costa \& Armandroff 1995).  
Therefore, the OH subsystem may also contain some accreted objects and 
is not a pure sample of GCs formed during the collapse 
of the protoGalaxy main body.  
NGC\,2419 and M\,5 might be some of these interlopers.
NGC\,2419 is the only BHB GC located beyond the Magellanic Clouds.  
While it has the same age as M\,92, an inner halo GC with 
similar metallicity (Harris et al.\,1997), 
it is quite difficult to imagine that this metal-poor cluster  
formed in the inner halo and then migrated into the far outer one.
The same problem stands for NGC\,5904 (M\,5), an outer halo BHB GC 
currently visiting the inner regions of the galactic halo 
(Dinescu et al.\,1999). 
Clearly, Zinn's classification needs refinements but nevertheless
constitutes a first step in understanding the 
different processes at work during the whole halo history.
Therefore, in what follows,  
we mainly rely on this BHB/RHB division. \\

To close this section, we note that some of the trends presented by GCs are 
also reproduced by field halo stars.
\begin{enumerate}
\item
According to Marquez \& Schuster (1994), field halo stars 
whose apogalactocentric 
distances are larger than $\simeq$\,10\,kpc form a younger group with a 
larger age dispersion than the inner part of the stellar halo.
\item
Majewski (1992) and Carney (1996) divide the stellar halo into a ``high
halo'' sample of stars ($\langle \vert Z_{Max} \vert \rangle \geq$ 
5\,kpc, where $Z_{Max}$ is the maximum height reached by the stars 
above the galactic plane) and a ``low halo''
($\langle \vert Z_{Max} \vert \rangle \leq$ 
5\,kpc).  The ``high halo'', which 
should be dominated by the accreted population, is  found 
in net retrograde rotation.  In contrast, the ``low halo'' is in 
prograde rotation.   
\item
Finally, Chiba (2000) notes that the stellar halo also includes subcomponents
characterized by different density distributions: the outer part 
of the halo ($D >$ 15\,kpc) appears to be nearly spherical, 
whereas the inner part exhibits a flattened distribution.  
\end{enumerate}
One interesting point is that the changes in these features 
occur more or less at the galactocentric distance
where the fraction of RHB clusters increases significantly. 
The stellar halo pattern is therefore consistent with the halo GC dichotomy.
This gives some support to the hypothesis of Jehin et al.\,(1999).
Following them, field halo stars
 were, at least partly, formed in GCs from which they 
escaped through various dynamical processes (disruption or evaporation).

\section{The metallicity gradient}
\subsection{Importance of the OH/YH division}
The whole galactic GCS exhibits a metallicity gradient interpreted 
as a disk-halo dichotomy, namely, the gradient 
is mainly driven by the high metallicity clusters (disk component: 
[Fe/H] $>-$ 0.8) located within $\simeq$\,8\,kpc from the galactic center
(Djorgovski \& Meylan 1994).  
The halo group itself ([Fe/H] $<-$ 0.8) presents no clear metallicity gradient
(see Fig.~\ref{FeHDWH}).
However, Sect.\,3 provides several arguments supporting a further 
meaningful division,
mainly based on a BHB/RHB (OH/YH) classification,
 of the halo system.  As a result,
the situation must be reconsidered. \\
Following Zinn (1993), the OH GCs are formed during the 
monolithic collapse of the protogalactic cloud while the 
more remote YH GCs are formed in fragments that escape the 
protogalactic collapse.  Later on, these fragments evolve into satellite 
systems, e.g. dwarf galaxies, bearing their own GCs into the 
galactic halo once they are accreted
by the Milky Way.
These dwarf galaxies GCs are therefore added to the 
{\it original galactic} GCs.  \\
According to the model exposed in Paper\,I,
the metallicity [Fe/H], induced by the self-enrichment process, is related 
to the hot protogalactic background pressure $P_h$
by (see Table \ref{tab}):
\begin{equation}
[{\rm Fe/H}] = 3.3 + 0.5 {\rm log} P_{h}\;.
\label{PFeH}
\end{equation}
Therefore, to a given pressure distribution of the hot 
protogalactic background, $P_h(D)$, corresponds a radial metallicity 
profile, [Fe/H]$(D)$.  GCs whose progenitor clouds were not
embedded in this pressure profile will corrupt the metallicity 
gradient if they are considered. \\
Since dwarf galaxy GCs were not born in the protoGalaxy, 
their progenitor clouds were not embedded in the same pressure 
profile as the galactic ones.  Consequently,
they must be rejected from 
any comparison between the theoretical results and the observational data. 

In  Fig.~\ref{FeHDWH}, the different types of GCs are marked by 
distinct symbols.  OH and YH clusters are respectively represented 
by open and full circles (lists of OH and YH GCs are provided in
Lee et al.\,1994 and Da Costa \& Armandroff\,1995).
The crosses label the 3 Sgr GCs with 
$\rm [Fe/H]<-\,0.8$.  The  open squares (``noHBR'' group) 
stand for the halo GCs
for which the HB morphology index is not given in Harris (1996),
probably because no color-magnitude diagrams precise enough to 
define the HB morphology index were available for these GCs.
Indeed, most of them are located in the vicinity of the galactic center and, 
therefore, there is no reason to think that they belong to the 
accreted component of the Galaxy.
The full triangle represents $\omega$\,Cen, the most peculiar galactic GC
(e.g. large mass, iron-peak element inhomogeneities 
in sharp contrast with the monometallicity observed in other GCs).
These peculiarities could be the result of the merger of two GCs 
(Jurcsik 1998) or of the accretion of a dwarf galaxy 
tidally stripped of its stellar envelope (Majewski 2000). 

Since the self-enrichment model applies to GCs formed in the proto-Milky 
Way all the progenitor clouds are embedded in a common pressure 
profile and we only consider the OH and noHBR groups.   
This subdivision of the halo GCs is important from the 
point of view of the existence or not of a metallicity gradient.
In Fig.~\ref{FeHDOH}, we show [Fe/H] vs log$D$, limited 
to the OH and noHBR groups.  By comparing Figs.~\ref{FeHDWH}
and \ref{FeHDOH}, we see that the GC metallicity appears to be 
more strongly correlated with log$D$ once the presumed accreted 
component is removed.  For the whole halo, the linear 
Pearson correlation coefficient is $-$0.3, corresponding to a 
probability of correlation of 99,85\,\%.  Considering only the GCs shown 
in  Fig.~\ref{FeHDOH}, this same coefficient improves to $-$0.49,  
corresponding to a probability of correlation larger than 99,999\,\%. \\  
Median values of log$D$ in four metallicity bins are given in Table 2
for the OH+noHBR groups  (the last column gives 
the number of clusters in each bin). 
The data in this table show a monotonic increase 
of $\langle$log$D\rangle$ with decreasing metallicity over the range 
[Fe/H]$\, \simeq -$0.8 to [Fe/H]$\, \simeq -$2.4.

\begin{table}[h]
  \caption{Metallicity vs log$D$ for the OH+noHBR groups.
  \label{median}}
  \begin{center}
    \leavevmode
    \footnotesize
    \begin{tabular}[h]{ccc}
      \hline \\[-5pt]
      [Fe/H] & $\langle$log$D\rangle$  &  n \\[+5pt]
      \hline \\[-5pt]
      $-1.2 \leq [{\rm Fe/H}] < -0.8 $ & 0.41 & 13  \\
      $-1.6 \leq [{\rm Fe/H}] < -1.2 $ & 0.64 & 28  \\
      $-2.0 \leq [{\rm Fe/H}] < -1.6 $ & 0.89 & 25  \\
      $-2.4 \leq [{\rm Fe/H}] < -2.0 $ & 1.02 &  9  \\
      \hline \\
      \end{tabular}
  \end{center}
\vspace{-0.7 cm}
\end{table}

On the average, the more remote GCs are more metal-poor.
This is in agreement with the self-enrichment model where the 
metallicity gradient is due to the decrease of the pressure exerted
by the hot protogalactic background on the PGCCs as the galactocentric 
distance increases (see Table\,\ref{tab}).  
To compare the theoretical metallicity gradient to 
the observational situation, one still needs an expression for the
pressure profile $P_{h}(D)$ (see Eq.~(\ref{PFeH})).

\subsection{Pressure profile of the hot protogalatic background}
Concerning this point, the situation is rather complex since there 
is obviously no agreement about the scaling of the $P_{h}(D)$ relation
in the literature.\\

The luminous components of galaxies form through the collapse of gas 
in gravitationally dominant halos of dark matter.
The density profile of such a halo  
is conveniently described as a singular isothermal sphere 
(White \& Kauffmann 1994):
\begin{equation}
\rho(D)=\frac{V_{c}^2}{4\pi G}\,\frac{1}{D^2}
\label{rhoD}
\end{equation}  
and
\begin{equation}
V_{c}^2=\frac{G M(D)}{D}=const.
\end{equation}  
In these equations, $V_c$ is the circular velocity of the gas in the 
dark matter potential well of the Galaxy and 
G is the gravitational constant.
Since we are mainly interested in the proto-Milky Way, 
the circular velocity $V_c$ is taken to be 220\,km\,s$^{-1}$
(Fall \& Rees 1985).
 
Two timescales are important in determining the further evolution 
of the gas component within this halo of dark matter : 
\begin{enumerate}
\item
the dynamical (or free-fall) time: 
\begin{equation}
\tau_{dyn}=\frac{D}{V_c},
\label{tdyn}
\end{equation}
\item
the radiative cooling timescale:
\begin{equation}
\tau_{cool}=\frac{3}{2} \frac{n_t k T}{n_i n_e \Lambda} 
\label{tcool}
\end{equation}
where $n_i$, $n_e$ and $n_t$ are respectively the ionic, electron 
and total number densities.
k is the Boltzmann constant, $T$ is the gas temperature, and 
$\Lambda$ is the cooling function (Sutherland \& Dopita 1993).
\end{enumerate}
When $\tau_{dyn} < \tau_{cool}$, 
the gas undergoes quasi-static contraction, but  
once $\tau_{cool} < \tau_{dyn}$, the gas cloud can no longer  
maintain itself in quasi-static equilibrium and the collapse 
proceeds on a free-fall timescale (e.g. Rees \& Ostriker 1977). \\

According to Fall \& Ress (1985), during the collapse of the 
protoGalaxy a two-phase structure,
namely cold clouds embedded in 
the remaining hot protogalactic background, grows in the collapsing gas.
The hot component is depleted by the condensation of the cold clouds until 
its cooling and dynamical timescales are comparable.  
The hot gas is expected to remain near the virial temperature of the halo 
described by Eq.~(\ref{rhoD}):
\begin{equation}
T_h=\frac{V_c^2}{2k} \mu_h m_H \simeq 1.7\times 10^6\, \rm K
\label{Tvir}
\end{equation}  
where $\mu_h$ is the mean molecular weight ($\sim 0.6$ for a 
primordial ionized plasma, with X $\simeq$ 0.76 and Y $\simeq$ 0.24).  
At this temperature, the value of the cooling function $\Lambda$ is
$4.8\times 10^{-24}$ erg\,cm$^3$\,s$^{-1}$ (Sutherland \& Dopita 1993).
Combining Eqs.~(\ref{tdyn} - \ref{Tvir}),
we have
\begin{equation}
P_h=5.3\times 10^{-10} D_{kpc}^{-1} \,\, \rm dyne\,cm^{-2}
\label{Ph1}
\end{equation}  
where $D_{kpc}$ 
is the galactocentric distance expressed in kpc. \\
The corresponding metallicity gradient is therefore 
(dashed line in Fig.~\ref{FeHDOH}):  
\begin{equation}
[{\rm Fe/H}] = -1.34 - 0.5\,{\rm log} D_{kpc}\,.
\label{grad1}
\end{equation}
In the Fall \& Rees model, the hot component of the gas in which
the PGCCs are embedded is at the virial temperature and
has a density such that $\tau_{cool} \simeq \tau_{dyn}$. 
The model proposed by Murray \& Lin (1992) is slightly different.  
It relies on the $\tau_{cool} \simeq \tau_{dyn}$ condition
and also on the hypothesis of hydrostatic equilibrium for the hot gas:
\begin{equation}
\frac{1}{\rho_h}\frac{{\rm d}P_h}{{\rm d}D} = -\frac{V_c^2}{D}.
\end{equation}
The peculiarity of their model is that the rotation curve of the collapsing
protoGalaxy is not assumed to be flat, but rather slightly decreasing
with galactocentric distance.
They argue that since the central regions, with the highest density, 
are the first to collapse, followed by regions from larger initial 
radii, the gravitational potential may be more centrally peaked 
than today.  They adopt $V_c(D) \propto D^{-1/4}$.  
It should be noted that the value of the exponent 
ensures that the temperature of the hot gas obeys to Eq.~(\ref{Tvir}) 
with $V_c$ and $T_h$ depending on $D$ (their Eq.~(3.9)).
In contrast with Fall \& Rees (1985),
their pressure profile for the hot protogalactic 
background scales as $D^{-2}$:   
\begin{equation}
P_{h} = 1.25\times 10^{-9} D_{kpc}^{-2}  \,\, \rm dyne\,cm^{-2}.
\label{Ph2}
\end{equation}
The corresponding metallicity gradient is therefore steeper 
(plain line in Fig.~\ref{FeHDOH}) than in Eq.~(\ref{grad1}):
\begin{equation}
[{\rm Fe/H}] = -1.15 - {\rm log} D_{kpc}\,.
\label{grad2}
\end{equation}

A third approach is suggested by Harris \& Pudritz (1994).
According to them, it is natural to expect that the hot and diffuse phase
of the protoGalaxy is at the virial temperature and in 
hydrostatic balance with the isothermal dark matter potential well.
Under these assumptions, the pressure distribution 
scales again as $D^{-2}$, $\tau_{dyn}$ and $\tau_{cool}$
scale respectively as $D$ (Eq.~(\ref{tdyn})) and as $D^{2}$ 
(Eq.~(\ref{tcool})).  There is thus a ``critical'' galactocentric 
distance $D_{crit}$ such that 
\begin{itemize}
\item[- ] $\tau_{cool} < \tau_{dyn}$ when $D < D_{crit}$: \\
no quasi-static equilibrium is possible; 
\item[- ] $\tau_{cool} \geq \tau_{dyn}$ when $D \geq D_{crit}$: \\
the gas can remain in hydrostatic equilibrium. 
\end{itemize}
In contrast to Fall \& Rees (1985) and Murray \& Lin (1992) models,
there is no equality between $\tau_{cool}$ and $\tau_{dyn}$ over 
the whole galactic halo and, therefore, this property can no longer 
be used to determine $P_h(D)$. However, it is interesting to note 
that the steady rise of the ratio $\tau_{cool}/ \tau_{dyn}$ 
with galactocentric distance is a property 
found in the halos of giant 
ellipticals, e.g. M\,87 (Fabricant et al.\,1980). \\

Murray \& Lin (1992) and Harris \& Pudritz (1994)
give the same scaling law for the pressure distribution
($P_h \propto D^{-2}$),
but not from the same hypotheses!  This illustrates the difficulty
to derive an expression for the
metallicity gradient from the self-enrichment model.  \\
In what follows we will use the results given by 
Eqs.~(\ref{grad1}) and (\ref{grad2}).

\subsection{Discussion}

\begin{figure}
\begin{center}
    \leavevmode
    \epsfxsize= 8.5 cm 
    \epsffile{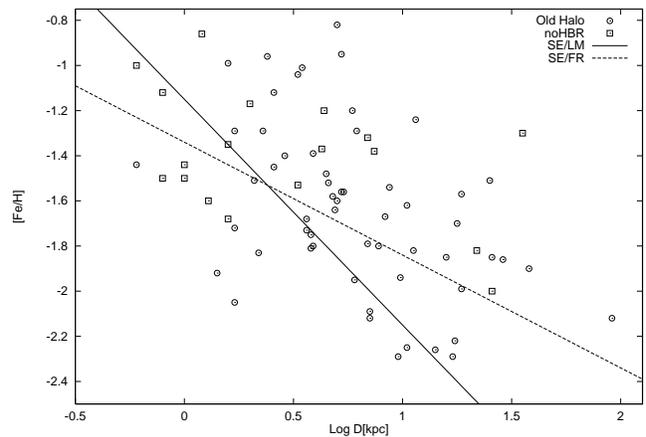}
  \end{center}
\caption{Comparison between 2 theoretical [Fe/H] vs log$D$ 
relationships and the metallicity radial distribution function of  
the OH+noHBR groups.  The plain and dashed curves
represent the self-enrichment model combined with, respectively,
the Murray \& Lin (1992) and the Fall \& Rees (1985) pressure 
profiles.  GC symbols are identical 
as in Fig.~\ref{FeHDWH} \label{FeHDOH}}
\end{figure}

In Fig.~\ref{FeHDOH}, we compare the results given by 
Eq.~(\ref{grad1}) (dashed line) and Eq.~(\ref{grad2}) 
(plain line) to the observational data.
The region of interest, namely where the bulk of the OH 
group is located, is the galactocentric range 1 to 30\,kpc.
The metallicity intervals predicted by Eqs.~(\ref{grad1}) and (\ref{grad2})
for these values of the galactocentric distance are respectively 
[$-$2\,dex, $-$1.35\,dex] and [$-$2.65\,dex, $-$1.15\,dex].  
The second model (self-enrichment model 
combined with $P_h \propto D^{-2}$, in this case the Murray \& Lin (1992) 
pressure distribution) is a better description for the 
observed metallicity range (see Table \ref{median}).  \\

The theoretical metallicity gradients corresponding to Eqs.~(\ref{grad1})
($P_h \propto D^{-1}$) and (\ref{grad2}) ($P_h \propto D^{-2}$)
are respectively  $\Delta [{\rm Fe/H}]/\Delta {\rm log}D = -0.5$\,dex
and $-$1\,dex.  
It is not straightforward to compare these results 
with the observational gradient because
of the rather high dispersion in the [Fe/H] vs log$D$ plot,
partly due to measurement errors in [Fe/H] and log$D$.  
The observed dispersion can also be explained by 
the GC orbital motions. 
Indeed, our model predicts a relation 
between the GC metallicities and the galactocentric distances of 
their {\it formation site}.  But the GCs were carried  
away from their formation sites through their orbital motions.
The initial radial distribution of globular cluster abundances 
has therefore been modified but we can only use the current one.
A least-squares fit to the OH+noHBR groups, which takes into
account the uncertainties in both coordinates (Press et al. 1992),
yields:
\begin{equation}
[{\rm Fe/H}]=(-1.35\pm 0.26){\rm log}D+(-0.64\pm 0.24).
\label{LS}
\end{equation}
If the noHBR clusters are not included, the slope of the 
least-squares fit is even steeper with a value of $-1.68\pm 0.41$. 
The errors on [Fe/H] are assumed to be $\pm$0.15\,dex (King~1999).
The errors on log$D$ are deduced from the comparison between the 
observational points and a ``classical'' least-squares fit to the
log$D$ vs [Fe/H] plot with $\sigma _{{\rm log}D}$=1.
This leads to an error of $\pm$0.4 in log$D$.  We caution that, by doing so,
the errors are assumed to be normally distributed.  This is 
probably not the case for log$D$, since the galactocentric distances of the
formation sites are replaced by the current galactocentric distances.
Therefore, the dispersion in log$D$ cannot
be attributed solely to measurement errors. The slope of Eq.~(\ref{LS})
is consistent with the one of Eq.~(\ref{grad2}) at the 1.3$\sigma$ level,
while it is different from the slope given by Eq.~(\ref{grad1})
at the 3.3$\sigma$ level.  As a conclusion, 
Eq.~(\ref{PFeH}) better describes the observed radial 
distribution of GC metallicities when it is combined with
a pressure profile scaling as $D^{-2}$ (Eq.~(\ref{grad2})) rather 
than $D^{-1}$ (Eq.~(\ref{grad1})).   
However, Eq.~(\ref{grad2}) somewhat underestimates the
mean observed metallicity at a given galactocentric distance.
This can be due to:
\begin{enumerate}
\item an underestimate by the self-enrichment model for the 
mass of metals ejected by the SNeII in the PGCCs (uncertainties in, e.g.,
the SNII yields, the high-mass stellar mass spectrum; see Paper I for a 
discussion of the second point); 
\item an underestimate of the hot protogalactic background pressure;
\item the possibility that GCs formed deeper in the galactic
potential well than their current location in the halo
(see Sect.\,5 for a related point). \\
\end{enumerate}  

In addition to the metallicity gradient, the model
presents a second interesting consequence.  
According to Sarajedini et al.\,(1997), once the search for 
an age-metallicity relationship is restricted to the Old Halo group, 
there is no evidence for such a relation.
This result is confirmed by the careful work carried out by  
Rosenberg et al.\,(1999).
Their conclusion is consistent with a single mean value for the age, 
independent of [Fe/H]\footnote{Again, the same pattern 
is observed for the field stars: 
there is no age-metallicity correlation for the FHS 
(Marquez \& Schuster 1994).} 
Clearly, a parameter other than the age is needed 
to explain the metallicity range observed in the OH/BHB group.  
The self-enrichment model provides us with such a parameter, 
namely the external pressure around the PGCCs.
Combined with a pressure distribution scaling as $D^{-2}$,
the model explains fairly  nicely
the galactic halo GC metallicities,
without any requirement for an age-metallicity relation
(in the usual sense of age decreasing with metallicity). 
The final proportion of heavy elements is mostly fixed by the 
pressure exerted by the medium surrounding the progenitor clouds. \\
If, in the future, the pressure distribution was proven 
to scale as $D^{-1}$ instead of
$D^{-2}$, then this would mean that our model 
underestimates/overestimates the amount of chemical
enrichment in the inner/outer regions of the halo. \\

Finally, the steady decline of the hot gas pressure with 
galactocentric distance should also lead to a gradient in the 
mass of the {\it PGCC} (see Table~\ref{tab}).  That such a mass gradient 
could still be observable today is not a certainty.  Indeed, the initial 
mass of the PGCCs undergoes changes due to various 
evolutionary processes: 
\begin{itemize}
\item[a. ] The formation of the second stellar generation:
there is no reason why this star formation episod would take place 
with the same star formation efficiency in each cloud.  
\item[b. ] Mass loss (evaporation): once they are formed, GCs 
undergo mass loss due to their interactions with the galactic 
gravitational field (Meylan 2000). 
\end{itemize}  
Evaporation and random star formation efficiency are sources of scatter 
in the relation between the PGCC and GC masses.  
In addition, mass determinations of GCs are still uncertain at 
least by a factor 2 (Meylan 2000).  
Therefore, and as pointed out by Brown (1993): ``Even if the Jeans 
mass did play a role in the formation of GCs, it is unlikely that a 
one-to-one relationship exists between the {\it proto-cluster cloud} mass 
and the {\it present cluster} mass.''  \\
GC absolute visual magnitudes $M_v$ could also be used to search 
for a relic of the mass gradient since
obtaining GC integrated luminosities is much easier than determining 
GC masses.  However, the absolute visual magnitude is not an accurate 
estimation of the GC mass because of the scatter introduced by the 
mass-to-light ratios in the mass-luminosity relation.  For instance, 
in the Pryor \& Meylan (1993) compilation, the mass-to-light ratios
of most of the OH GCs range from 1.1 to 3.8.  Despite these various 
uncertainties, Brown (1993) finds a steady decline in $M_v$ for 
$D<15$\,kpc in a plot of average $M_v$ versus average log$D$ for the 
galactic halo clusters.  Therefore, his conclusion further supports 
the hypothesis of an origin of galactic halo GCs in clouds having
the Jeans mass.

\section{Were GCs formed near their perigalacticon?}

\begin{figure}
\begin{center}
    \leavevmode
    \epsfxsize= 8.5 cm 
    \epsffile{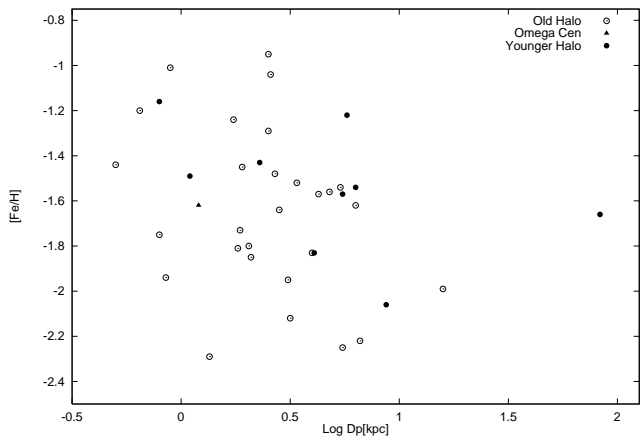}  
  \end{center}
\caption{Same as in Fig.~\ref{FeHDWH} but the GC metallicities are plotted 
vs the perigalactocentric distances \label{FeHDpWH}}
\end{figure}

The initial halo metallicity gradient
was severely altered by the accretion of GCs formed in dwarf and/or 
irregular galaxies.  The genuine Milky Way GCs moving away 
from their formation sites is another cause of alteration. 
It would be very useful to know the galactocentric distances at which the 
GCs were formed, in order to better evaluate the validity of 
Eqs.~(\ref{grad1}) and (\ref{grad2}).  \\
In Sect.\,4.3, it was suggested that GCs might have formed 
deeper in the galactic halo than their current location.
However, we caution that this result relies on the self-enrichment 
model {\it and} on the assumed pressure distribution.
Therefore, this outcome cumulates uncertainties from both
and it certainly does not stand on a firm support.
Even so, previous papers already suggest that the correlation
between [Fe/H] and log$D_p$, where $D_p$ is the perigalactocentric 
distance of the GCs, may be stronger than between [Fe/H] and log$D$. \\

\begin{figure}
\begin{center}
    \leavevmode
    \epsfxsize= 8.5 cm 
    \epsffile{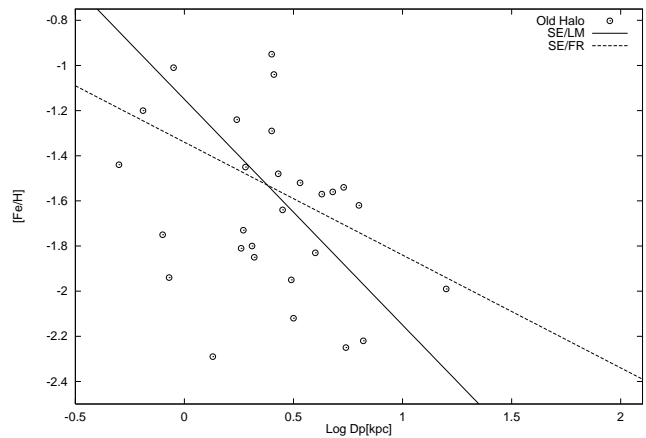}
  \end{center}
\caption{Same as in Fig.~\ref{FeHDOH} but the GC metallicities are plotted 
vs the perigalactocentric distances \label{FeHDpOH}}
\end{figure}

Freeman \& Norris (1981) notice ``a clear gradient in the 
[Fe/H]-$D_p$ plane''.  
According to them, this may be a hint that the clusters 
did form near perigalacticon.  Van den Bergh (1995) also notes
that the GC metallicity correlates somewhat more strongly with $D_p$
than it does with the present GC galactocentric distances.
Nevertheless, the method used to derive these perigalactic distances 
is somewhat problematic.  It assumes that the current GC tidal
radius is mostly set by the galactic tidal field at the closest 
approach of the cluster to the galactic center, namely the 
perigalacticon (King 1962).  As a result, 
tidal radii of well observed GCs are used to estimate 
their perigalactic distances.  By so doing, the GC internal processes,
especially the two-body relaxation, are neglected (Meziane \& Colin 1996).
These internal processes lead to a replenishment of the outer regions 
of the cluster between two perigalactic passages and therefore modify  
the outer radius set at the perigalacticon.  
Thus, it is dangerous to rely on the current outer radius of GCs to derive
their perigalactocentric distances.
In order to avoid this problem, perigalactic distances 
derived from the computation of GC orbits (Dinescu et al.\,1999)
are used in Figs.~\ref{FeHDpWH} (all GCs with known perigalactic distances)
and \ref{FeHDpOH} (OH GCs).  Another cluster, NGC 6522 (Terndrup et al.\,1998),
is added to Dinescu's list. 
The sample is by far smaller than in Van den Bergh (1995).
This is due to the necessity to know the proper motions in order 
to compute the GC orbits in a given galactic potential. \\ 
The (log$D_p$, [Fe/H]) plot (Fig.~\ref{FeHDpOH}) does not appear tighter 
than the (log$D$, [Fe/H]) one (Fig.~\ref{FeHDOH}).
The linear Pearson correlation coefficient in Fig.~\ref{FeHDpOH} 
is only $-$0.31, corresponding to a probability of correlation of 
the order of 90\,\%.
In contrast, considering the (log$D$, [Fe/H]) plot for the 
same sample of GCs, the correlation coefficient 
is still $-$0.43, corresponding to a probability of correlation of 
96\,\%. 
The orbits only provide evidence that the more metal-rich 
halo clusters ([Fe/H] $> -$1.4) are concentrated towards the galactic 
center ($D_p <$ 3\,kpc).  Therefore, these new data do not really 
confirm the suggestion made by  Freeman \& Norris (1981).  
New perigalactic distances would be helpful to give a definitive 
answer to the existence of a link between the perigalactocentric 
distances and the metallicities of galactic halo GCs.

\section{Conclusions}

The self-enrichment model of galactic halo GCs (see Paper\,I for 
a detailed description) has been compared
to the observational situation and the conclusions are as follow:

\begin{enumerate}

\item The final metallicity induced by the self-enrichment process
depends on $P_h$, the pressure exerted on the PGCC by the hot 
protogalactic background medium.  
The result is in agreement with galactic halo GC metallicities.

\item There is a range of halo GC metallicities due to the variations 
in $P_h$ with the galactocentric distance.  Considering 
the clusters located between 1 and 30\,kpc, the ranges of metallicities 
[Fe/H] are
0.75 and 1.50\,dex when the self-enrichment model is combined with
$P_h \propto D^{-1}$ and $P_h \propto D^{-2}$ respectively.
The second solution is in better agreement with the observed 
metallicity range. 
Moreover, when the Murray \& Lin (1992) pressure distribution  is used,
the theoretical metallicity interval over the galactocentric range 
$1 \leq D_{kpc} \leq 30$ is $-2.65\leq [{\rm Fe/H}] \leq -1.15$.
This result is in nice agreement with the observations
in the galactic halo GCs.
  
\item Because of the expected decrease in $P_h$ with increasing 
galactocentric distance, the model induces a metallicity gradient 
throughout the galactic halo.
Such a metallicity gradient is indeed observed once the GCs
suspected to have been accreted by the Milky Way are removed.
These clusters were born in fragments which evaded the initial 
protogalactic collapse and have experienced their own chemical
evolution.  The division between the galactic and accreted 
components of the halo is mainly based on the BHB/RHB (OH/YH)
classification introduced by Zinn (1993). 
Indeed, it seems likely that the halo GCs consist 
of clusters with more than one origin.
Since the accreted clusters did not take part in the formation and 
early evolution of the galactic halo, 
their progenitor clouds did not share the same external 
pressure distribution as the genuine galactic proto-GCs.
Therefore, they should not be taken into account in 
the self-enrichment model. \\
Again, the observed radial distribution of GC abundances favours
a background pressure profile scaling as $D^{-2}$ rather than $D^{-1}$. \\
However, it should be noted that the scatter of the data about the 
model lines in Fig.~\ref{FeHDOH} (and also in Fig.~\ref{FeHDpOH}) 
exceeds the observational 
uncertainties.  As already stated, this can be due to the GC 
orbital motions which carry them away from their formation sites.
Furthermore, it is certainly an oversimplification to consider 
that $P_h$ is the only parameter determining the GC metallicity.
Other parameters must interfere (e.g. stellar 
mass range, SNII yields, ...). 
\item In our model,
no age-metallicity relation is required to explain the different 
GC metallicities. Actually, there is no compelling 
evidence for an age-metallicity relationship 
among halo GCs (Buonanno et al.\,1998), and especially once the sample is 
limited to the BHB/OH group (Sarajedini et al.\,1997).
In this group, all GCs are coeval according to 
Rosenberg et al.\,(1999). 
This is in agreement with our self-enrichment model where we see 
an enhanced chemical enrichment with decreasing galactocentric distance
rather with time. 
\end{enumerate}

\begin{acknowledgements}

This research was supported by contracts P\^ole d'Attraction 
Interuniversitaire P4/05 (SSTC, Belgium) and
FRFC F6/15-OL-F63 (FNRS, Belgium).
  
\end{acknowledgements}


\begin{thebibliography}{}

\bibitem[1993]{Brown}
Brown J.H., 1993, In: Smith G.H., Brodie J.P. (eds) ASP Conference 
Series Volume 48, The globular clusters-galaxy connection, p.~766 
\bibitem[1991]{Brown}
Brown J.H., Burkert A., Truran J.W. 1991, ApJ 376, 115 
\bibitem[1995]{Brown}
Brown J.H., Burkert A., Truran J.W. 1995, ApJ 440, 666
\bibitem[1993]{Buonanno}
Buonanno R., Corsi C.E., Fusi Pecci F., Richer H.B., Fahlman G.G. 
1993, AJ 105, 184
\bibitem[1995a]{Buonanno}
Buonanno R., Corsi C.E., Fusi Pecci F., Richer H.B., Fahlman G.G. 
1995a, AJ 109, 650
\bibitem[1998]{Buonanno}
Buonanno R., Corsi C.E., Pulone L., Fusi Pecci F., Bellazini M. 1998,
A\&A 333, 505
\bibitem[1990]{Carignan}
Carignan C., Beaulieu S., Freeman K. 1990, AJ 99, 178
\bibitem[1996]{Carney}
Carney B.W., Laird J.B., Latham D.W., Aguilar L.A. 1996, AJ 112, 668
\bibitem[1986]{Cayrel}          
Cayrel R. 1986, A\&A 168, 81 
\bibitem[1996]{Chaboyer}
Chaboyer B., Demarque P., Sarajedini A. 1996, ApJ 459, 558 
\bibitem[2000]{Chiba}
Chiba M., 2000, In: A.~Weiss, T.~Abel and V.~Hill (eds.) 
Proceedings of the 2$^{nd}$ ESO/MPA Conference, 
The First Stars, p.\,77 (Springer)
\bibitem[1999]{Cote}
C\^ot\'e P. 1999, AJ 118, 406
\bibitem[1993]{Da Costa}
Da Costa G.S., 1993, In: Smith G.H., Brodie J.P. (eds) ASP Conference 
Series Volume 48, The globular clusters-galaxy connection, p.~363 
\bibitem[1995]{Da Costa}
Da Costa G.S., Armandroff T.E. 1995, AJ 109, 2533
\bibitem[1999]{Dinescu}
Dinescu D.I., Girard T.M., Van Altena W.F. 1999, AJ 117, 1792 
\bibitem[1994]{Djorgovski}
Djorgovski S., Meylan G. 1994, AJ 108, 1292
\bibitem[1962]{Eggen}
Eggen O.J., Lynden-Bell D., Sandage A. 1962, ApJ 136, 748
\bibitem[1988]{Elson}
Elson R.A.W., Fall S.M. 1988, AJ 96, 1383
\bibitem[1980]{Fabricant}
Fabricant D., Lecar M., Gorenstein P. 1980, ApJ 241, 552
\bibitem[1985]{Fall}
Fall S.M., Rees M.J. 1985, ApJ 298, 18 
\bibitem[1995]{Ferraro}
Ferraro I., Ferraro F.R., Fusi Pecci F., Corsi C.E., Buonanno R. 1995,
MNRAS 275, 1057 
\bibitem[1981]{Freeman}
Freeman K.C., Norris J. 1981 ARA\&A 19, 319
\bibitem[1995]{Fusi Pecci}
Fusi Pecci F., Bellazini M., Cacciari C., Ferraro F.R. 1995, AJ 110, 1664
\bibitem[1994]{Harris}
Harris W.E., Pudritz R.E. 1994, ApJ 429, 177 
\bibitem[1996]{Harris}
Harris W.E. 1996, AJ, 112, 1487 
\bibitem[1997]{Harris}
Harris W.E., Bell R.A., Vandenberg D.A., Bolte M., Stetson P.B., 
Hesser J.E., van den Bergh S., Bond H.E., Fahlman G.G., Richer H.B.
1997, AJ 114, 1030
\bibitem[1987]{Hartwick}
Hartwick F.D.A., 1987, In: The Galaxy; Proceedings of the NATO 
Advanced Study Institute, Cambridge, England, Dordrecht, 
D. Reidel Publishing Co., p.~281-290.
\bibitem[1997]{Ibata}
Ibata R.A., Wyse F.G.A., Gilmore G., Irwin M.J., Suntzeff N.B. 1997,
AJ 113, 634 
\bibitem[1995]{Irwin}
Irwin M.J., Demers S., Kunkel W.E. 1995, ApJ 453, 21
\bibitem[Jehin]{1999}
Jehin E., Magain P., Neuforge C., Noels A., Parmentier G., 
A. Thoul 1999, A\&A 341, 241
\bibitem[1999]{Johnston}
Johnston K. V., Majewski S. R., Siegel M. H., Reid I. N. and Kunkel W. E.
1999, AJ 118, 1719
\bibitem[1998]{Jurcsik}
Jurcsik J. 1998, ApJ 506, L113
\bibitem[1962]{King}
King I.R. 1962, AJ 67, 471
\bibitem[1999]{King}
King I.R., 1999, In: Martinez Roger C., P\'erez Fournon I., Sanchez F. (eds) 
Cambridge University Press, Globular Clusters, p.~1
\bibitem[1992]{Larson}
Larson R.B., 1992, In: Tenorio-Tagle G., Prieto M., Sanchez F. (eds) 
Cambridge University Press, 
Star Formation in Stellar System, p.~143
\bibitem[1993]{Lee}
Lee Y.W., 1993, In: Smith G.H., Brodie J.P. (eds) ASP Conference 
Series Volume 48, The globular clusters-galaxy connection, p.~142
\bibitem[1994]{Lee}
Lee Y.W., Demarque P., Zinn R. 1994, ApJ 423, 248  
\bibitem[1992]{Lin}
Lin D.N.C., Richer H.B. 1992, ApJ 388, L57
\bibitem[1996]{McLaughlin}
McLaughlin D.E., Pudritz R.E. 1996, ApJ 469, 194
\bibitem[1992]{Majewski}
Majewski S.R. 1992, ApJS 78, 87
\bibitem[1994]{Majewski}
Majewski S.R. 1994, ApJ 431, L17
\bibitem[2000]{Majewski}
Majewski S.R., 2000, In: A. Noels, P. Magain, D. Caro, 
E. Jehin, G. Parmentier, A. Thoul (eds.) 35$^{th}$ Li\`ege 
International Astrophysics Colloquium, The galactic halo: 
from globular clusters to field stars, p~619
\bibitem[1994]{Marquez}
Marquez A., Schuster W.J. 1994, A\&AS 108, 341
\bibitem[1998]{Mateo}
Mateo M. 1998, ARA\&A 36, 435
\bibitem[1997]{Meylan}
Meylan G., Heggie D.C. 1997, A\&AR 8, 1
\bibitem[2000]{Meylan}
Meylan G., 2000, In: A. Noels, P. Magain, D. Caro, 
E. Jehin, G. Parmentier, A. Thoul (eds.) 35$^{th}$ Li\`ege 
International Astrophysics Colloquium, The galactic halo: 
from globular clusters to field stars, p~543
\bibitem[1996]{Meziane}
Meziane K., Colin J. 1996, A\&A 306, 747
\bibitem[1996]{Mighell}
Mighell K.J., Rich R.M., Shara M., Fall S.M. 1996, AJ 111, 2314
\bibitem[1995]{Minniti}
Minniti D. 1995, AJ 109, 1663
\bibitem[1992]{Murray}
Murray S.D., Lin D.N.C. 1992, ApJ 400, 265  
\bibitem[1997]{Odenkirchen}
Odenkirchen M., Brosche P., Geffert M., Tucholke H.-J. 1997
New Astronomy 2, 477
\bibitem[1999]{Parmentier}
Parmentier G., Jehin E., Magain P., Neuforge C., Noels A., Thoul A.A. 1999,
A\&A 352, 138  
\bibitem[1992]{Press}
Press W.H., Teukolsky S.A., Vetterling W.T. and Flannery B.P. 1992, 
Numerical Recipes (2nd ed.; Cambridge Univ. Press)
\bibitem[1993]{Pryor}
Pryor C., Meylan G. 1993, In: S.G. Djorgovski, G. Meylan (eds)
ASP Conference 
Series Volume 50, Structure and Dynamics of globular clusters, p.~370  
\bibitem[1977]{Rees}
Rees M.J., Ostriker J.P. 1977, MNRAS 179, 541
\bibitem[1999]{Rosenberg}
Rosenberg A., Saviane I., Piotto G., Aparicio  A. 1999, AJ 118, 2306
\bibitem[1996]{Sarajedini}
Sarajedini A., Geisler D. 1996, AJ 112, 2013
\bibitem[1997]{Sarajedini}
Sarajedini A. 1997, AJ 113, 682
\bibitem[1997]{Sarajedini et al.}
Sarajedini A., Chaboyer B., Demarque P. 1997, PASP 109, 1321 
\bibitem[1978]{Searle}
Searle L., Zinn R. 1978, ApJ 225, 357
\bibitem[1989]{Stetson}
Stetson P.B., VandenBerg D.A., Bolte M., Hesser J.E., Smith G.H. 
1989, AJ 97, 1360
\bibitem[1996]{Stetson}
Stetson P.B., VandenBergh D.A., Bolte M. 1996, PASP 108, 560 
\bibitem[1999]{Stetson}
Stetson P.B., Bolte M., Harris W.E., Hesser J.E., van den Bergh S., 
VandenBergh D.A., Bell R.A., Johnson J.A., Bond H.E., Fullton L.K.,
Fahlman G.G., Richer H.B. 1999, AJ 117, 247  
\bibitem[1993]{Sutherland}
Sutherland R.S., Dopita M.A. 1993, ApJS 88, 253
\bibitem[1998]{Terndrup}
Terndrup D.M., Popowski P., Gould A., Rich R.M., Sadler E.M. 
1998 AJ 115, 1476.
\bibitem[1995]{Van Den Bergh}
Van Den Bergh S. 1995, AJ 110, 1171
\bibitem[1994]{White}
White S.D.M., Kauffmann G., 1994, In: Munoz-Tunon C., Sanchez F. (eds) 
Cambridge University Press, 
The Formation and Evolution of Galaxies, p.~471
\bibitem[1985]{Zinn}
Zinn R. 1985, ApJ 293, 424
\bibitem[1993]{Zinn}
Zinn R., 1993, In: Smith G.H., Brodie J.P. (eds) ASP Conference 
Series Volume 48, The globular clusters-galaxy connection, p.~38




\end{thebibliography}
\end{document}